**Manuscript Category: Review**

**Resting-state functional connectivity-based biomarkers and functional MRI-based neurofeedback for psychiatric disorders: a challenge for developing theranostic biomarkers**

Takashi Yamada[1, 2*], Ryu-ichiro Hashimoto[1, 2, 3, 4*], Noriaki Yahata[1, 5, 6], Naho Ichikawa[7], Yujiro Yoshihara[8], Yasumasa Okamoto[7], Nobumasa Kato[2], Hidehiko Takahashi[8], Mitsuo Kawato[1†]

* These two authors equally contributed to the present work

†Corresponding author

**Affiliations**

[1]Department of Decoded Neurofeedback, ATR Brain Information Communication Research Laboratory Group, Advanced Telecommunications Research Institute International, Kyoto, Japan

[2]Medical Institute of Developmental Disabilities Research, Showa University, Tokyo, Japan

[3]Department of Language Sciences, Graduate School of Humanities, Tokyo Metropolitan University, Tokyo, Japan

[4]Research Center for Language, Brain and Genetics, Tokyo Metropolitan University, Tokyo, Japan

[5]Department of Molecular Imaging and Theranostics, National Institute of Radiological Sciences, National Institutes for Quantum and Radiological Science and Technology, Chiba, Japan

[6]Department of Youth Mental Health, Graduate School of Medicine, The University of Tokyo, Tokyo, Japan

[7]Department of Psychiatry and Neurosciences, Graduate School of Biomedical Sciences, Hiroshima University, Hiroshima, Japan

[8]Department of Psychiatry, Kyoto University Graduate School of Medicine, Kyoto, Japan

**Correspondence**

Mitsuo Kawato

2-2-2 Hikaridai, Seika-cho, Sorakugun, Kyoto, Japan

Phone: +81-774-95-1111

Email: kawato@atr.jp

Fax: +81-774-95-1236

**Word count**

Abstract: 236; Manuscript: 5976

**Abstract**

Psychiatric research has been hampered by an explanatory gap between psychiatric symptoms and their neural underpinnings, which has resulted in poor treatment outcomes. This situation has prompted us to shift from symptom-based diagnosis to data-driven diagnosis, aiming to redefine psychiatric disorders as disorders of neural circuitry. Promising candidates for data-driven diagnosis include resting-state functional connectivity MRI (rs-fcMRI)-based biomarkers. Although biomarkers have been developed with the aim of diagnosing patients and predicting the efficacy of therapy, the focus has shifted to the identification of biomarkers that represent therapeutic targets, which would allow for more personalized treatment approaches. This type of biomarker (i.e., "theranostic biomarker") is expected to elucidate the disease mechanism of psychiatric conditions and to offer an individualized neural circuit-based therapeutic target based on the neural cause of a condition. To this end, researchers have developed rs-fcMRI-based biomarkers and investigated a causal relationship between potential biomarkers and disease-specific behavior using functional MRI (fMRI)-based neurofeedback on functional connectivity. In this review, we introduce recent approach for creating a theranostic biomarker, which consists mainly of two parts: (i) developing an rs-fcMRI-based biomarker that can predict diagnosis and/or
3

symptoms with high accuracy, and (ii) the introduction of a proof-of-concept study investigating the relationship between normalizing the biomarker and symptom changes using fMRI-based neurofeedback. In parallel with the introduction of recent studies, we review rs-fcMRI-based biomarker and fMRI-based neurofeedback, focusing on the technological improvements and limitations associated with clinical use.

**Introduction**

Although great advancements in psychiatric research have been made in recent years, an explanatory gap between phenomenological entities and neurobiological underpinnings remains (Montague et al., 2012). This gap has prevented precise diagnosis and dramatic improvements in treatment outcomes in the field of clinical psychiatry (Insel and Cuthbert, 2015). Our lack of understanding of the disease mechanisms is reflected by the fact that the two world-wide standard psychiatric diagnosis systems—the Diagnostic and Statistical Manual of Mental Disorders (DSM) (American Psychiatric Association, 2013) and International Classification of Diseases (ICD) (World Health Organization, 1990)— adopt symptom-based approaches, in which underlying biological substrates are not taken into consideration (Insel and Cuthbert, 2009), except in the case of dementia. Consequently, these symptom-based diagnostic systems may artificially draw distinctions among conditions that actually share common biological etiologies, and therefore, may fail to provide effective biology-based treatments directed toward specific pathogenic processes associated with these conditions (Owen, 2014). Therefore, the recent initiative of research domain criteria (RDoC) has proposed an important paradigm shift from the conventional symptom-based categories to data-driven dimensional approaches based on observable behaviors and neurobiological measures (Insel and Cuthbert, 2009), with the aim of eliminating the gap between disease-related behaviors and neurobiological substrates. In this review, we first explain that brain functional measures provided by fMRI, particularly the resting-state functional connectivity (rs-fc) MRI, play crucial roles for the development of "biomarkers" that provide dimensions along which various psychiatric disorders could be defined. Then, we illustrate the potential and power of the fMRI- and biomarker-based neurofeedback methods in the treatments of disease-related behaviors. By illustrating the development of rs-fcMRI-based biomarkers and fMRI-based neurofeedback, we claim that these two lines of new research converge in filling the abovementioned gap.

**Towards development of theranostic biomarker for psychiatric disorder**

Recent psychiatric neuroimaging research has begun to bridge the explanatory gap by redefining psychiatric disorders as disorders of neural circuitry (Insel and Cuthbert, 2015). Indeed, rs-fcMRI represents a promising platform for identifying affected neural circuitry. Traditionally, alterations in neural circuitry have been studied by examining brain activation and/or functional connectivity (FC) during specific task conditions using a limited number of participants. However, more recent whole-brain rs-fcMRI studies have applied state-of-the-art machine learning algorithms to "big data" in order to identify brain features that predict the diagnostic status and/or severity of psychiatric disease (Clementz et al., 2016; Yahata et al., 2016 and 2017; Arbabshirani et al., 2017). Since brain features are identified in a data-driven manner, this approach is free of potential biases that may derive from explicit hypotheses regarding affected brain regions, FCs, or functions. Furthermore, relative to task-based fMRI, rs-fcMRI is more suited to the clinical investigation of patients and young children who have difficulty in performing tasks (Poldrack and Farah, 2015). In addition, previous studies have shown that resting brain signals generate highly structured spatiotemporal patterns that correspond well to those observed when performing tasks (Smith et al., 2009; Laird et al., 2011). These signals predict brain activation evoked by several kinds of tasks at the level of individual participant (Tavor et al., 2016). These studies indicate that rs-fcMRI data may include abundant information regarding individual characteristics and that this data can therefore be used as a substitute for task-based fMRI data. Together with the relative simplicity and low variability in data acquisition setups, rs-fcMRI serves as the platform by which large amounts of clinical data can be analyzed to develop appropriate machine-learning algorithms.

Many studies along this line of psychiatric research share the goal of identifying biological measures of altered neural circuitry that represent "biomarkers" for psychiatric disorders (Perlis, 2011; Abi-Dargham and Horga, 2016). Indeed, to date, a number of structural and functional MRI studies have claimed to have identified such "biomarkers" for various psychiatric disorders (Fan et al., 2008; Sun et al., 2009; Kim et al., 2010; Sui et al., 2015; Ivleva et al., 2016; Kambeitz et al., 2016; Drysdale et al., 2017; Li et al., 2017). However, the significance of these identified biomarkers varies greatly depending on study aims and designs. Here, we propose that so-called "biomarkers" should be categorized into the following four types: (i) biomarker "candidates" that correlate with diagnosis in a sample pool, (ii) those that generalize over the studied samples and therefore predict diagnosis of a disease of interest in a general population, (iii) those that predict the effect of a therapy (i.e., surrogate endpoint), and (iv) those that correspond to the disease mechanism and may therefore be regarded as therapeutic targets. Biomarker types (i) and (ii) are similar in that the measure simply represents a correlation with the disease status, though they are critically different regarding whether the scope of the biomarker is limited to the sample dataset (i) or has the capacity to generalize over the disease of interest in general, beyond the sample (ii). In this sense, only biomarker types (ii)-(iv) qualify as true "biomarkers" (Abi-Dargham and Horga, 2016). While biomarker types (ii) and (iii) can be used as an auxiliary test in clinical practice and are expected to provide important information regarding the diagnosis and treatment strategy, the clinical importance of these two types of biomarkers does not necessarily indicate that these measures account for the disease mechanism. For instance, low-density lipoprotein (LDL) cholesterol has been used as a surrogate marker for a clinically meaningful endpoint for the heart disease. However, lowering LDL cholesterol does not necessarily lead to the prevention of heart disease at the individual level because the density of LDL cholesterol may not be associated with the underlying cause of heart disease (Albert, 2011). In contrast, when the biomarker corresponds to elements of the disease mechanism as in type (iv), its significance is two-fold: as a measure of diagnostic status and/or the severity of symptoms, and as a therapeutic target. Therefore, we refer to this type of biomarker as a "theranostic biomarker". The development of such theranostic biomarkers will result in breakthroughs not only in basic biological research but also in clinical psychiatry practice, providing patients with individually tailored therapeutic targets and allowing for the elimination of unnecessary treatments and adverse effects (Ahn, 2016).

Hereafter, we introduce several recent studies that have suggested that some rs-fcMRI-based biomarkers satisfy prerequisites for type (ii), (iii), and even (iv) biomarkers for major psychiatric disorders. That is, these biomarkers may explain elements of disease mechanisms and identify a therapeutic target for a range of neuromodulation interventions, including neuropharmacology, repetitive transcranial magnetic stimulation (rTMS), and neurofeedback. In order to strictly verify that the rs-fcMRI-based biomarker represents the disease mechanism, the following three levels of evidence are necessary: (A) The rs-fcMRI-based biomarker predicts diagnostic status and/or the severity of symptoms with high accuracy for the general population of a disease of interest; (B) normalization of the biomarker via neuromodulation interventions leads to the alleviation of symptoms; and (C) alterations of neural circuits that are represented by the biomarker are caused by a whole range of known risk factors for the disease of interest, including genes, molecules, cells, circuits, cognition, behavior, and the physical and social environments. However, it is very difficult to provide the third level of evidence due to the limited datasets obtained in human studies. Consequently, we discuss the first two levels of evidence to examine whether rs-fcMRI-based biomarkers are capable of acting as theranostic biomarkers for psychiatric disorders. Concretely, we first introduce the development of the rs-fcMRI-based biomarkers for autism spectrum disorder (ASD), major depressive disorder (MDD), schizophrenia (SCZ), and obsessive compulsive disorder (OCD)—which utilized state-of-the-art machine learning algorithms that achieved high classification accuracy and generalized well for independent validation cohorts. Secondly, we introduce the preliminary results of recent proof-of-concept studies that have examined whether the normalization of rs-fcMRI-based biomarker can be achieved via fMRI-based neurofeedback on FC, and whether such normalization leads to the improvement of symptoms in depression. In addition, we also refer to technical difficulties in the development of rs-fcMRI-based biomarkers and fMRI-based neurofeedback and discuss recent advances in overcoming these challenges.

**Rs-fcMRI-based biomarker for psychiatric disorder**

*The importance of rs-fcMRI-based biomarkers for psychiatric disorders*

The brain generates highly structured spatiotemporal patterns even in the absence of explicit task execution (i.e., under resting-state conditions) (Smith et al., 2009; Laird et al., 2011). This finding suggests that rich information may be decoded by applying machine learning algorithms to rs-fcMRI data in the individual brain. Indeed, a series of studies have successfully used such algorithms to predict various characteristics in healthy individuals, including age (Dosenbach et al., 2010), intelligence (Smith et al., 2015), working memory (Yamashita et al., 2015), and sustained attention (Rosenberg et al., 2016). Based on these successful applications, a growing number of studies have sought to develop rs-fcMRI-based biomarkers for various psychiatric disorders (Arbabshirani et al., 2017), such as ASD (Anderson et al., 2011), MDD (Drysdale et al., 2017), SCZ (Kaufmann et al., 2015), and ADHD (Deshpande et al., 2015).

*The generalization ability of rs-fcMRI-based biomarkers*

A number of rs-fcMRI-based biomarker studies have claimed high accuracy in discrimination between individuals with a disease of interest and healthy controls (HC) for most major psychiatric disorders. However, to date, no such biomarkers have been identified for use in routine clinical practice. Aside from issues related to economic and practical feasibility in clinical settings, one major issue with previously developed biomarkers is that accuracy in discrimination of the biomarker is validated only for a single sample cohort that is shared with the training of the biomarker. Therefore, the generalizability of the biomarker is usually untested beyond the sample dataset, and highly accurate discrimination is likely to fail when that biomarker is applied to an independent cohort. More specifically, if the developed biomarker is fitted to noise structures that are specific to the training dataset (e.g., demographic distributions and measurement conditions such as the type of MRI scan protocol), the prediction is inflated for the training data but catastrophic to the independent validation dataset, which does not contain the same noise structure (Whelan and Garavan, 2014; Huys et al., 2016; Yahata et al., 2017).

In order to develop clinically meaningful rs-fcMRI-based biomarkers, it is necessary to prove the generalizability of the biomarker using independent datasets as validation cohorts. For this step to be successful, the development of optimal machine-learning algorithms that alleviate over-fitting to the noise structures of the training data is critical. Such over-fitting often occurs when a large number of parameters are included relative to the number of participants, and when the model does not sufficiently remove the effect of nuisance variables that are included in training dataset (Whelan and Garavan, 2014; Yahata et al., 2017). Therefore, for the model to be reliable, the number of parameters in the model should be reduced based on the number of participants, and the brain features that reflect disease-related factors (e.g. diagnostic status and symptom severity) should be extracted after removing the data-specific noise structure. In the following section, we review the development of rs-fcMRI-based biomarkers that satisfy the aforementioned conditions for ASD (Yahata et al., 2016), MDD (Ichikawa et al., 2017), SCZ (Yoshihara et al., 2017), and OCD (Takagi et al., 2017). In illustrating these cases, we show that the overfitting problem was successfully alleviated by the development of novel machine-learning algorithms, which resulted in the identification of a small number of altered functional connections (FCs) capable of discriminating between individuals with a specific medical condition of interest and healthy or typically developed controls (HC or TD). The resultant biomarker has achieved high accuracy for a discovery cohort (i.e., training data), together with good generalizability for independent validation cohorts (i.e., test data).

*The rs-fcMRI-based biomarker for ASD*

Although it is generally believed that abnormal FCs may underlie ASD (Menon, 2011), whether such abnormalities involve under-connectivity, over-connectivity, or distance-dependent alterations (Just et al., 2012; Supekar et al, 2013; Long et al., 2016) remains unknown. Several research groups have attempted to solve this problem by developing rs-fcMRI-based biomarkers. However, none of these biomarkers has been validated in an independent cohort (Anderson et al., 2011). One study that attempted to validate the generalizability of the biomarker observed poor performance below chance in an independent cohort (Yoshihara et al. 2011). Among these unsatisfactory attempts to develop an rs-fcMRI biomarker for ASD, Yahata et al. (2016), aimed to achieve a desired level of generalizability by controlling the two causes of over-fitting: the number of parameters in the model and the interference of nuisance variables. Specifically, they developed a unique combination of machine-learning algorithms of L1 regularized sparse canonical correlation analysis (L1-SCCA) followed by sparse logistic regression (SLR; Yamashita et al., 2008). Briefly, in this algorithm, L1-SCCA was applied to extract FC features associated with diagnostic labels (e.g., ASD or TD), while removing FC features associated with nuisance variables (e.g., age, sex, medication, scan protocol). Then, sparse estimation performed by L1-SCCA and SLR reduced the number of explanatory variables (i.e., FCs) in the biomarker. Therefore, the combination of L1-SCCA and SLR is highly suited for controlling the aforementioned two causes of over-fitting inherent to machine-learning studies using multicenter rs-fcMRI data.

Yahata et al. (2016) applied this novel machine learning algorithm to rs-fcMRI data from 74 high-functioning adults with ASD and 107 TD adults obtained from three different sites in Japan. FC data in each individual were analyzed as a correlation matrix representing the Pearson correlation values for 9,730 pairs of time-series data extracted from 140 regions in the sulci-based anatomical atlas (extended Brainvisa Sulci Atlas; Perrot et al., 2011). Using the correlation matrices of 181 individuals as inputs, the machine learning algorithm of L1-SCCA and SLR generated a classifier consisted of only 16 FCs (0.2% of all FCs) that distinguished between ASDs and TD with a high accuracy of 85% and an area under the curve (AUC) of 0.93 (Figure 1*a*).

Because the biomarker for ASD was developed using Japanese datasets only, it must be validated using independent cohort datasets, which, in this study, were collected in countries with different cultural and ethnic backgrounds than those in Japan. Therefore, the US ABIDE dataset was selected as an independent cohort (Di Martino et al., 2014), which consisted of 44 high-functioning adults with ASD and 44 demographically matched TD controls. Indeed, the biomarker developed in Japan generalized well and exhibited a high classification accuracy of 75% (AUC=0.76) (Figure 1*b*). To our knowledge, this is the first study to demonstrate high generalizability for an independent cohort across cultures and ethnicities. The generalizability of the biomarker was further confirmed in a second independent cohort collected in Japan (Accuracy = 70%, AUC = 0.77). Lastly, the selected FCs used in the developed biomarker predicted with high accuracy not only diagnostic status but also the severity of communication problems, based on communication domain scores of the Autism Diagnostic Observation Schedule (ADOS) (Lord et al., 2000) ($r = 0.44$, $p < 0.001$). These results indicate that, with the proper use of machine-learning algorithms for controlling over-fitting, we are able to develop a reliable biomarker from the rs-fcMRI data that predicts ASD with high accuracy.

Further, we applied this ASD biomarker to other psychiatric disorders (SCZ, ADHD, and MDD) to investigate whether the selected FCs could discriminate patients with these psychiatric disorders from healthy controls. That is, we aimed to determine whether the biomarker is specific to ASD diagnosis. Our results indicated that this biomarker could not significantly differentiate individuals with ADHD or MDD from their respective controls (ADHD: AUC = 0.57, $p$ = 0.65, MDD: AUC = 0.48, $p$ = 0.83), although moderate differentiation of those with SCZ was observed (AUC = 0.65, $p$ = 0.012) (Figure 1$c$). This modest generalizability of the constructed ASD biomarker only to SCZ indicates that the weighted summation of the extracted FCs for the biomarker may reflect the extent of "ASD-ness", or more precisely liability of ASD, as individuals throughout any population—including those with other psychiatric conditions—may possess ASD-like traits. This speculation is biologically plausible considering the evidence that ASD is closer to SCZ than ADHD and MDD in terms of genetic, behavioral, and neuroimaging findings (King and Lord, 2011; Cross-Disorder Group of the Psychiatric Genomics Consortium, 2013).

**Figure 1**

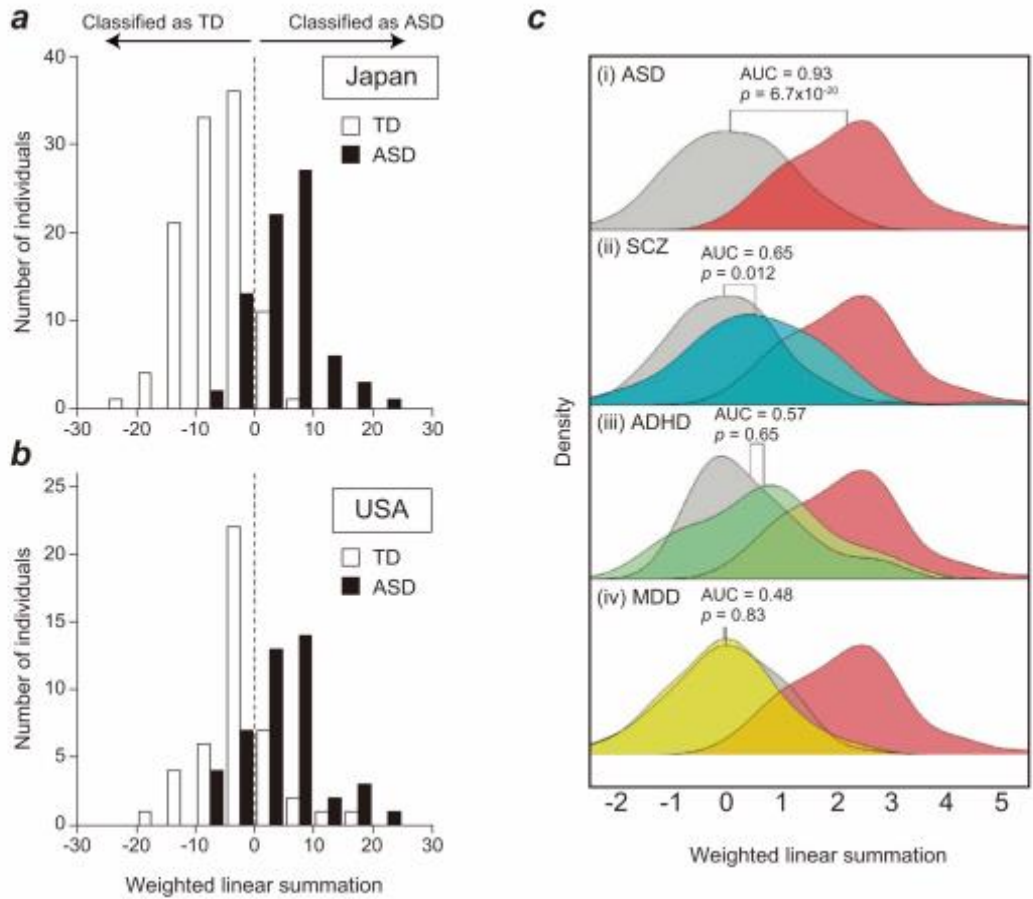

*The generalizable rs-fcMRI-based biomarkers for depression, schizophrenia, and obsessive compulsive disorder*

The classifier construction algorithm developed in Yahata et al. (2016) was applied to patients with melancholic depression, SCZ, and OCD. Ichikawa et al. (2017) restricted the training samples to individuals with melancholic depression (N = 66) and demographically matched healthy controls with Beck Depression Inventory-II (BDI-II) scores ≤ 10 (Beck et al., 1996) (N = 66) to avoid issues in the heterogeneity of depression (CONVERGE consortium, 2015). The resulting classifier identified 12 diagnosis-specific FCs (out of 9,316 connections) and achieved high discriminability (AUC = 0.77, Accuracy = 70%) in the training dataset and high generalizability (AUC = 0.62, Accuracy = 65%) to the independent validation cohort (i.e., test data), which included individuals with melancholic depression (N = 11) and healthy controls (N = 40). (These healthy controls were matched based on BDI score to those in the training dataset.)

A biomarker for SCZ (AUC = 0.83, Accuracy = 76%) using a Japanese training dataset consisting of patients in the chronic stage (duration of illness = 12.8 ± 7.8 years), scanned at two imaging sites in Japan, was also developed (Yujiro Yoshihara, personal communication, March 30, 2017). The SCZ biomarker also worked well for two independent validation cohorts of patients in the chronic stage (USA site 1: duration of illness = 14.2 ± 11.5 years, AUC = 0.75, Accuracy = 70%, EU site: duration of illness = 5.9 ± 5.8 years, AUC = 0.66, Accuracy = 61%), but not for a test dataset of patients experiencing their first episode (USA site 2: duration of illness = 16.8 ± 9.6 months, AUC = 0.42, Accuracy = 45%). These findings suggest that each stage of SCZ is associated with specific pathological processes manifested as differentially altered FCs, in accordance with the findings of previous studies (Anticevic et al., 2015; Li et al., 2016).

For the discrimination of OCD, Takagi et al. (2017) adopted the methodology for developing rs-fcMRI based biomarkers described by Yahata et al (2016) in conjunction with principal component analysis for feature selection of FCs, partly because a small training data set from a single imaging site was available. All principal components were analyzed using the aforementioned algorithm (L1-SCCA and SLR). The biomarker for the training dataset ($N_{ocd}$ = 52, $N_{hc}$ = 56) exhibited high accuracy (AUC: 0.81, Accuracy: 73%), while that for the test dataset ($N_{ocd}$ = 18, $N_{hc}$ = 10) exhibited high generalizability (AUC: 0.70).

**Neurofeedback**

*fMRI-based neurofeedback*

Neurofeedback is an auxiliary technique for self-regulating the neural activity that underpins specific behaviors or symptoms by providing participants with real-time feedback that represents the current activity state of the neural activity of interest (Sitaram et al., 2016; Thibault et al., 2016). In contrast to other neuromodulation methods that rely on externally applied factors (e.g. electromagnetic field and pharmacological agents), neurofeedback is a method of internally (either volitionally or conditionally) regulating neural activity. As such, this method provides a means to aid participants to learn to induce brain activity toward a desired pattern of neural activity relying only on participants' own endogenous factors. Among several neuroimaging modalities including electroencephalography (EEG) and functional near-infrared spectroscopy (fNIRS), fMRI-based neurofeedback has attracted considerable attention for its potential as a novel method of therapeutic treatment in clinical neuroscience (Fovet et al., 2015; Morimoto and Kawato, 2015). To date, several studies have applied fMRI-based neurofeedback methods to psychiatric patients by training them to up-regulate or down-regulate the level of activation in single or multiple regions-of-interest (ROI) (Sitaram et al., 2016). Although this method has proven effective for some psychiatric conditions (Linden et al., 2012; Scheinost et al., 2013; Sitaram et al., 2014; Zilverstand et al., 2017), recent advances in fMRI data acquisition and analysis have enabled us to evaluate fMRI signals with greater spatial and temporal precision in real-time. This technological advance has yielded two novel lines of fMRI-based neurofeedback: decoded neurofeedback (DecNef) and functional-connectivity-based neurofeedback (FCNef). DecNef is a novel method of controlling distributed activity patterns of multiple voxels within a circumscribed ROI (Shibata et al., 2011) (Figure 2), thereby greatly increasing spatial precision over the previous method of regulation of localized activity in coarsely defined ROIs. In contrast, FCNef allows for the modulation of spatiotemporal activation patterns across multiple ROIs (Koush et al., 2013; Koush et al., 2015) as well as intrinsic functional networks (Megumi et al., 2015) (Figure 3). In the following sections, we review recent successful applications of DecNef and FCNef, which, in our view, have potential as novel interventions for the treatment of psychiatric disorders via powerful regulation of neural activity (Yanagisawa et al., 2016; Sakai, 2016).

**Figure 2**

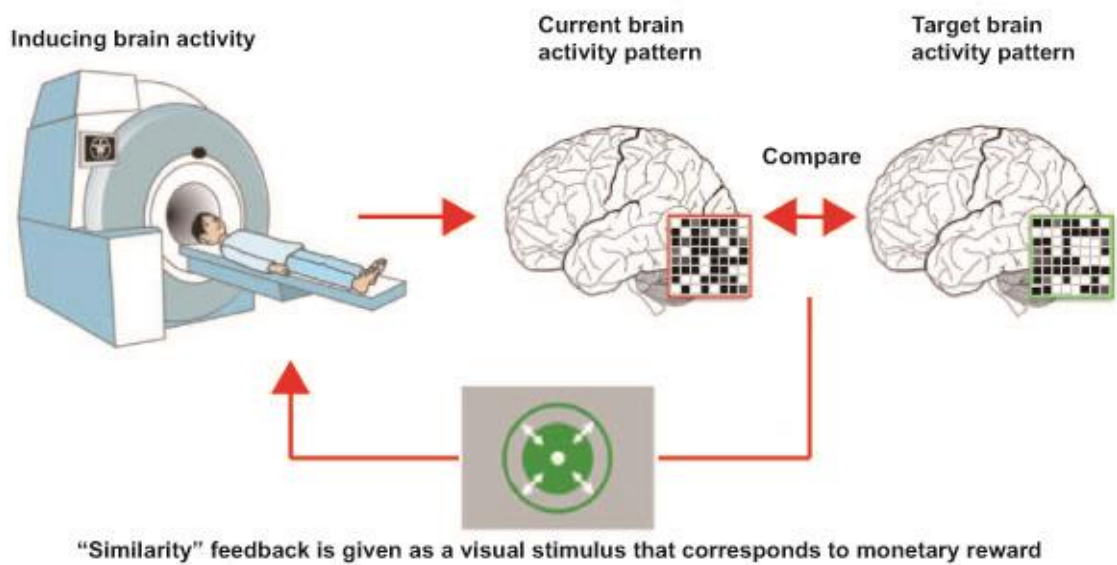

**Figure 3**

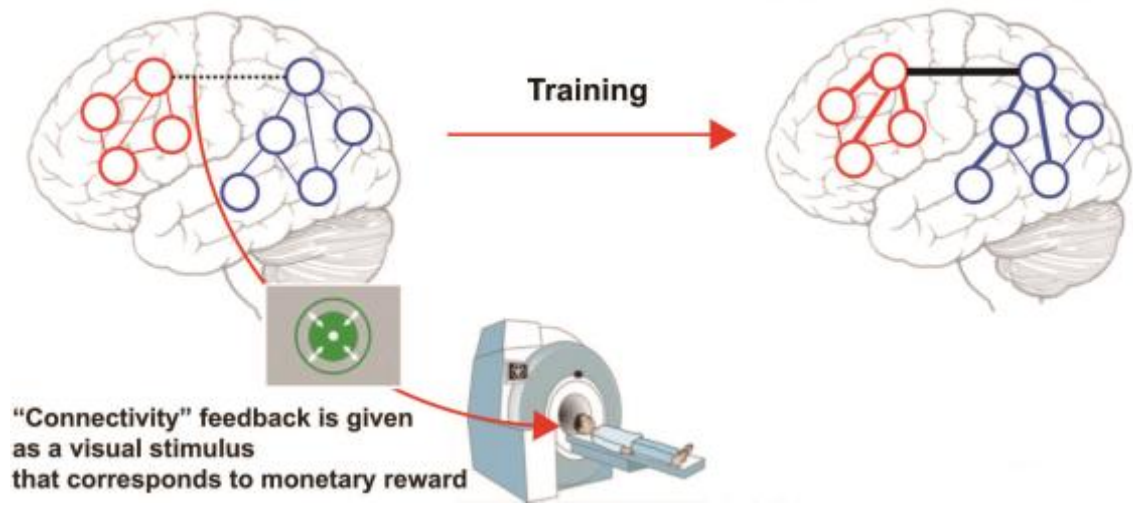

*Decoded neurofeedback (DecNef)*

DecNef is a novel neurofeedback method in which participants learn to induce specific activation patterns of multiple voxels in a given brain region. This neurofeedback method is based on recently developed fMRI decoding techniques (Kamitani and Tong, 2005), which allow researchers to infer the mental experiences and states of participants via analysis of multi-voxel patterns of activation. DecNef is thus based on the assumption that, by determining a particular multi-voxel pattern as a target pattern that corresponds to a specific mental experience or state, experimenters can calculate the similarity between the target and the current multi-voxel pattern online and return it to participants as a feedback score. DecNef studies are usually composed of three types of experiments: (i) pre- and post-behavioral tests, (ii) fMRI decoder construction, and (iii) DecNef training (Shibata et al., 2011, 2016; Amano et al., 2016; Koizmi et al., 2016; Cortese et al., 2016 and 2017). The pre- and post-behavioral tests examine whether the DecNef method can alter the target behavior in the desired direction (e.g., whether DecNef can enhance visual perceptual learning). In the fMRI decoder construction stage, multi-voxel patterns of activation corresponding to the target mental experience or state (e.g., visual perception of gratings with orientation of 45 degrees) are identified via a stimulus-driven and task-based fMRI experiment. In the DecNef training stage, participants learn to induce the decoded multi-voxel pattern of activation matched with the target mental experience or state through neurofeedback. In the following paragraph, we mainly review three DecNef studies, in which researchers aimed to alter three types of behavior: 1) visual perceptual learning, 2) meta-cognition, and 3) fear response.

Shibata et al. (2011) demonstrated that the DecNef method may aid participants in inducing target spatiotemporal patterns of activation in the primary visual cortex corresponding to a specific orientation of Gabor patches, without the presentation of a matched stimulus. Furthermore, the authors observed that this method resulted in visual perceptual learning specific to the target orientation. These results and characteristics indicate that DecNef has the potential to induce sufficient neuro-plasticity for perceptual learning in the adult early visual cortex with high selectivity.

Based on the findings of this seminal study, DecNef has now been extended to the associative learning (Amano et al., 2016), face preference (Shibata et al., 2016), meta-cognition (Cortese et al., 2016) and fear extinction paradigms (Koizumi et al., 2016). In the following paragraphs, we focus on the latter two paradigms, which are thought to be related to the etiology and/or pathogenesis of psychiatric disorders. Cortese et al. (2016) selected confidence ratings as the target behavior during a two-choice dot-motion discrimination task in a cross-over DecNef study. (That is, the same participant performed two types of DecNef in a random order, separated by a 1-week interval: one aimed at increasing confidence ratings and one aimed at decreasing confidence ratings.) The authors reported that DecNef could be used to modulate activity in fronto-parietal regions to produce bi-directional alterations in confidence ratings without affecting task accuracy. Such a result would indicate that confidence is well decoded in higher cognitive areas, and that DecNef can be used to bi-directionally alter this meta-cognition-related behavior.

Koizumi et al. (2016) investigated whether the DecNef paradigm could be applied to fear extinction. In this experiment, pre-behavioral testing included fear-conditioning, in which fear responses were induced by pairing two kinds of colored circles (target fear-conditioned stimulus CS+ and control CS+) with electric shocks. Participants then underwent 3 days of DecNef training in which rewards were paired with the multi-voxel patterns of activity in V1/V2 matched to the target CS+. After the training, fear responses as assessed by skin conductance response were selectively reduced for the target CS+ but not for control CS+. We emphasize that this counter-conditioning occurred even though participants were not explicitly exposed to any fear-related stimuli, but rather implicitly exposed to the neural activity patterns matched with the target CS+.

The results of the latter two types of experiments—which may reflect alterations in meta-cognitive processes and the Pavlovian conditioned fear response—suggest that DecNef may be useful as an adjunctive therapy for psychiatric disorders, as meta-cognition and fear are closely related to behavioral changes associated with mental illness (David et al., 2012). In particular, counter-conditioning DecNef may benefit patients with fear-related disorders such as phobias and post-traumatic stress disorder, as explicit exposure to traumatic situations (e.g., prolonged exposure therapy) may be too difficult for some patients (Schnurr et al., 2007).

*Functional-connectivity-based neurofeedback (FCNef)*

The scope of fMRI-based neurofeedback now extends beyond controlling activation levels or patterns within ROIs to include regulation of functional connectivity between brain regions. Kim et al. (2015) demonstrated that the combination of ROI- and functional connectivity-based neurofeedback for heavy smokers aided participants in inducing increased ROI activation and functional connectivity, which is accompanied by reduced cravings for nicotine. Another FCNef study used dynamic causal modeling to enhance the flow of information from the dorsomedial prefrontal cortex to the amygdala—the putative neural circuit associated with the cognitive control of emotions—successfully reducing state anxiety (Koush et al., 2015).

These results indicate that specific regulation of FCs can indeed be achieved using FCNef, in turn leading to desired changes in behavior and function. Although such findings suggest that FCNef may represent a novel treatment method for psychiatric disorders, it remains unclear for how long FCNef-induced connectivity changes are retained. This issue was addressed by another recent study in which FC between the lateral parietal and primary motor areas, which were negatively correlated prior to training, was enhanced via a 4-day FC-based neurofeedback training protocol (Megumi et al., 2015). This increase in functional connectivity during the training period resulted in positive alteration of the rs-fcMRI between the default-mode and motor/visuo-spatial networks, which include the two ROIs, respectively. Intriguingly, this effect lasted for more than 2 months after the training. These results indicate that FCNef may be capable of inducing robust and long-lasting plasticity in target FCs, which is clinically significant for the treatment of psychiatric disorders. Yamashita et al. (2017) further demonstrated that FCNef induced bidirectional changes in behavior by changing the sign of a neurofeedback signal. Our hypothesis is as follows: If an rs-fcMRI-based biomarker capable of discriminating between individuals with a psychiatric condition and healthy controls with high accuracy can be developed, successful normalization of the individual's own FC pattern using FCNef would lead to a reduction in psychiatric symptoms.

*Unique characteristics of DecNef and FCNef*

DecNef and the latter two FCNef (Megumi et al., 2015; Yamashita et al., 2017) studies have the following three unique characteristics; (i) implicitness, (ii) monetary reward, and (iii) spatially limited influence. First, no verbal instruction regarding any explicit strategy was given to participants, and no participant became aware of how feedback was increased or the mechanisms underlying the neurofeedback experiment. Second, monetary reward was given to participants in proportion to the success of voxel pattern or functional connectivity induction. Third, induced information by DecNef and FCNef was largely constrained in the brain region. As for (i), no participant adopted a rational cognitive strategy that was fitted to the respective experimental designs, as revealed in post-experiment interviews (Shibata et al., 2011 and 2016; Megumi et al., 2015; Amano et al., 2016; Koizumi et al., 2016; Cortese et al., 2016 and 2017). When an efficient cognitive strategy is unavailable, desired brain activation can be reinforced by providing contingent feedback and/or rewards, rather than by the cognitive strategy itself. Therefore, reinforcement learning—or more specifically, neural operant conditioning—may well explain the training mechanism of DecNef and FCNef. However, in order to develop a clinically useful neurofeedback paradigm, it is necessary to compare the effect of various instruction, feedback, and reward conditions in future studies.

*Neurofeedback and pharmacology*

When applying these fMRI-based neurofeedback methods to individuals with psychiatric disorders, it is necessary to consider the relationship between neurofeedback and pharmacology, as many patients with such conditions are prescribed psychotropic agents. One previous study examined the interaction between neurofeedback and pharmacology in rodents (Ishikawa et al., 2014), reporting that mice were capable of inducing hippocampal neuronal activity through a neurofeedback operant conditioning method using lateral hypothalamus stimulation as a contingent reward. The authors further demonstrated that an NMDA receptor antagonist and dopamine D1 receptor antagonist abolished the neural operant conditioning. In contrast, depression model mice conditioned by forced swimming failed to induce target neural activity, although this ability was recovered following treatment with an antidepressant. These results suggest that psychotropic agents may act to either enhance or diminish the effect of neurofeedback in humans, indicating that combined treatment with neurofeedback and pharmacological agents may improve clinical outcomes.

**Neurofeedback therapy based on neuroimaging biomarkers**

Based on the promising results of the aforementioned studies regarding rs-fcMRI-based biomarkers and FCNef, several proof-of-concept studies have examined the potential efficacy of functional-connectivity-based neurofeedback in the treatment of patients with MDD and ASD (Hashimoto 2013; Kawato 2013; Yahata et al., 2016 and 2017). The most recent study consists of (i) rs-fcMRI-based biomarker construction (see *rs-fcMRI-based biomarker for psychiatric disorder*) and (ii) normalization (i.e., FCs consisted of the biomarker) using an FCNef protocol. The protocol for normalization of target FCs was determined in large part based on a previous study (Megumi et al., 2015). Briefly, the FCNef training was held over 4 successive days. In each trial during the training, participants were instructed to manipulate brain activity to increase as much as possible the size of a green disc in the display, which represented the degree of target FC normalization. The following paragraphs discuss the preliminary findings of recent proof-of-concept experiments in individuals with MDD and ASD. These studies were approved by the ethical committee of Kyoto University and Showa University, respectively. All volunteers gave written informed consent prior to the study, in accordance with the Declaration of Helsinki.

In the study of MDD, we selected FC between the left dorsolateral prefrontal cortex (DLPFC) and left precuneus/posterior cingulate cortex (Precun/PCC) as a target for FCNef, based on the following steps. First, we constructed two types of rs-fcMRI-based biomarker, one for predicting diagnosis (i.e., depression or healthy) (Ichikawa et al., 2017) and the other for predicting the severity of depressive symptoms (i.e., the score of BDI) (Yamashita et al., 2015). The target FC was defined as that included in both types of biomarkers. The identified FC was consistent with the findings of previous studies that have demonstrated an imbalance in anti-correlation between the default-mode and fronto-parietal networks as a neural correlate for MDD (Kaiser et al., 2015; Northoff, 2016; Rayner et al., 2016). During neurofeedback training, participants aimed to decrease the correlation of the target FC. In the most recent study, FCNef has been conducted for three individuals with MDD and seven individuals with subclinical depression. Participants with average BDI-II scores >10 at two different time points prior to training were categorized into the subclinical depression group. Figure 4*a* shows the neurofeedback scores of all three patients with MDD on each day of training. Scores exhibited an upward trend across the 4 days of training, and this was confirmed using a multiple regression model that included two explanatory variables (i.e., each training day and subject) and one response variable (i.e., neurofeedback scores), showing significant positive effect of the training day (95% confidence interval [CI] 1.9-9.1 of the coefficient). Post-hoc *t*-tests revealed that scores for all three participants were significantly higher on the last day than on the first ($t = 4.01$, $p < 0.001$). These results consistently demonstrated that participants learned to induce negative correlation for the target FC throughout the training. Furthermore, all three patients exhibited decreased scores on the Hamilton Depression Rating Scale (HDRS) (Hamilton, 1980), which represents the severity of depressive symptoms, after the training (Figure 4*b*). Similar to those of the depression group, neurofeedback scores also tended to increase over the training period among the seven individuals with subclinical depression (Figure 4*c*). One-way analysis of variance (ANOVA) revealed a significant main effect of training day ($p = 0.011$), while post-hoc paired *t*-tests revealed that neurofeedback scores were significantly higher on the last day of training than on the first ($p = 0.0046$). A tendency for reduced BDI scores following training was also observed ($p = 0.07$). Furthermore, five of the seven participants also reduced the target rs-fcMRI in the normal direction, and the change in rs-fcMRI between pre- and post-FCNef was significantly correlated with that of BDI score ($r = 0.87$, $p = 0.011$) (Figure 4*d*). These results indicated the possibility that the target FC between the left DLPFC and left precuneus/PCC may be a theranostic biomarker for depression, as more than half of the participants decreased the target rs-fc in accordance with BDI score through our FCNef training. Taken together, these findings suggest that our therapeutic package for depression can be used to detect potential theranostic biomarkers and ameliorate depressive symptoms using circuit-specific fMRI-based neurofeedback.

The integration of the FCNef technique with disease-specific rs-fcMRI-based biomarkers may also aid in the development of a novel therapeutic treatment for ASD. Using the highly accurate rs-fcMRI-based biomarker for ASD (Yahata et al., 2016), we conducted a proof-of-concept study of this approach in which a small number of adults with high-functioning ASD underwent 4 to 5 successive days of FC-neurofeedback training (Hashimoto 2013; Yahata et al., 2016 and 2017). In contrast to MDD, multiple FCs included in the ASD biomarker were selected as targets of intervention. Although the results are still preliminary and several aspects of the protocol must be refined, we observed steady improvement in feedback scores throughout the training in some participants. This observation indicates that some individuals with ASD are indeed capable of learning to change their altered FC patterns in the direction toward the typically-developed pattern, even in adulthood. Furthermore, we observed cases in which the neurofeedback training had a long-lasting impact on the FC pattern during the resting-state. The outputs of the rs-fcMRI-based biomarker closely approached the neurotypical level not only during the training sessions but also more than three weeks after the training, whereas, prior to training, the biomarker outputs had been invariably ASD-like. A typical case is shown in Figure 5. We acknowledge that even more robust changes in rs-fcMRI may be required to induce behavioral changes that may significantly improve patient quality of life. Furthermore, several difficulties may exist that are specific to ASD, such as the altered sensitivity to reward (Dichter et al., 2012; Kohls et al., 2013). However, recent preliminary results suggest that real-time FCNef guided by the disease-specific rs-fcMRI-based biomarker may provide a foundation for the development of a novel neuro-circuitry-based therapy, particularly for conditions in which the effects of standard interventions are very limited, such as ASD.

**Figure 4**

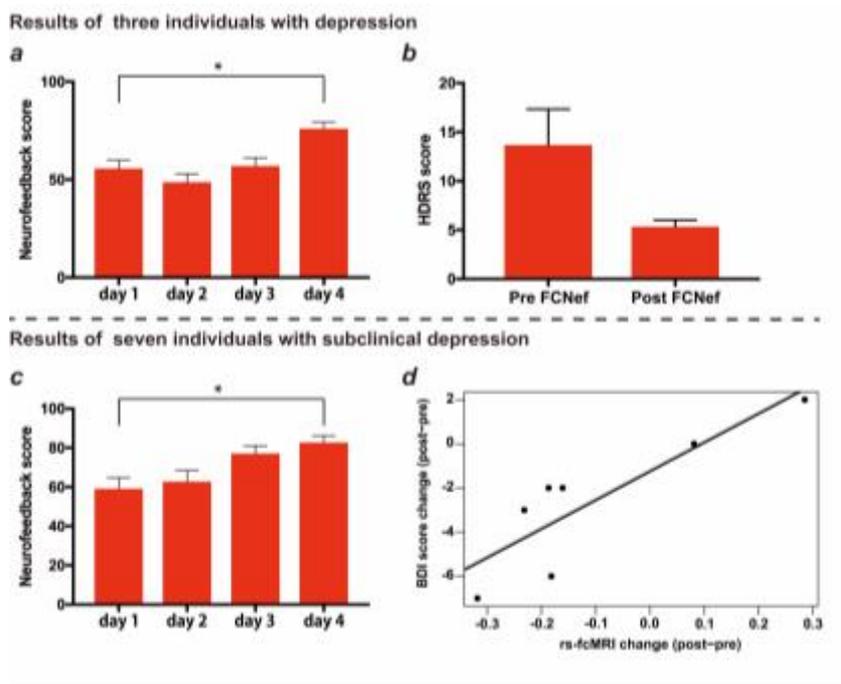

**Figure 5**

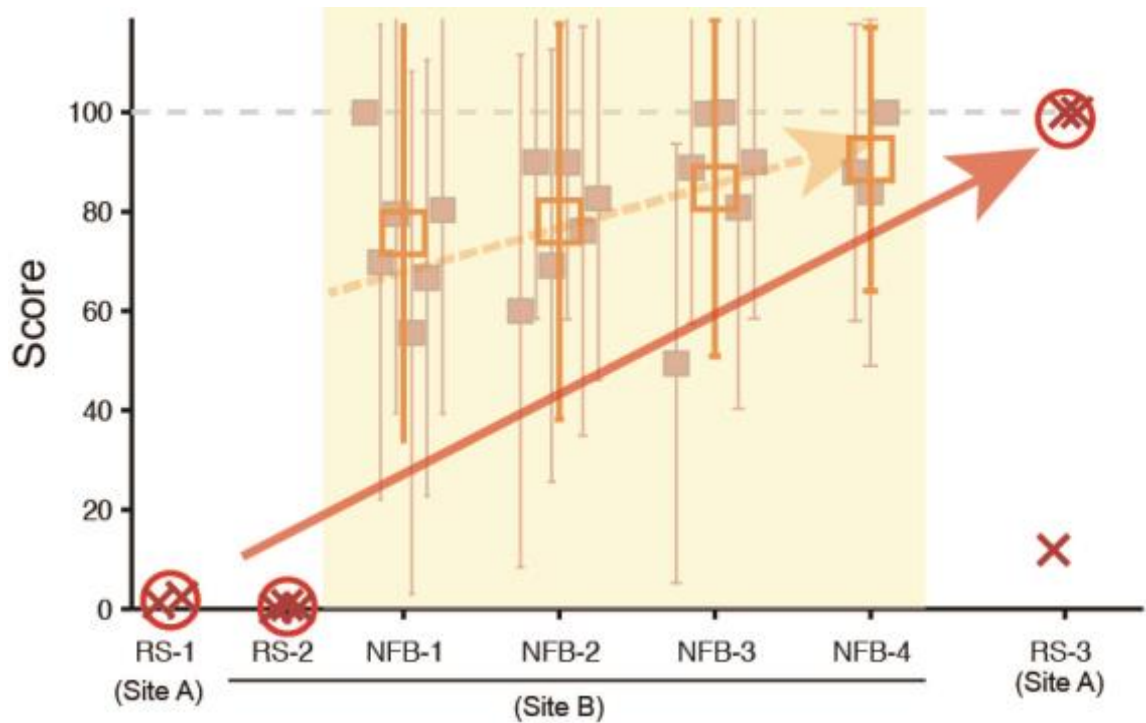

**Conclusion**

In this review, we discussed recent progress in computational psychiatric studies and the findings of our research program focusing on rs-fcMRI-based biomarkers and fMRI-based neurofeedback (DecNef project 2017). While not utilized in clinical psychiatry at present, these approaches have the potential to change the conventional method of symptom-based diagnosis to a data-driven method, allowing for more precise treatment with psychotropic agents and circuit-specific therapies such as neurofeedback and rTMS. Furthermore, the combination of rs-fcMRI-based biomarkers and FCNef may allow for the simultaneous diagnosis and treatment of psychiatric disorders, thus establishing a theranostic biomarker, which has yet to be achieved in clinical psychiatry.

**Funding**

This research is conducted as the "Application of DecNef for development of diagnostic and cure system for mental disorders and construction of clinical application bases" of the Strategic Research Program for Brain Sciences from Japan Agency for Medical Research and development, AMED. T.Y. was also partially supported by the Japan Society for the Promotion of Science (JSPS) Grant-in-Aid for Scientific Research (C) (16K10236) and grants from SENSHIN Medical Research Foundation.

**Acknowledgments**

We thank T. Kochiyama for the assistance in MRI data analysis. We thank A. Yamashita and S. Hayasaka for the assistance in planning neurofeedback paradigm.

Montague PR, Dolan RJ, Friston KJ, Dayan P (2012) Computational psychiatry. Trends Cogn Sci 16:72-80.

Morimoto J, Kawato M (2015) Creating the brain and interacting with the brain: an integrated approach to understanding the brain. J R Soc Interface 12:20141250.

Northoff G (2016) Spatiotemporal psychopathology I: No rest for the brain's resting state activity in depression? Spatiotemporal psychopathology of depressive symptoms. J Affect Disord 190:854-866.

Owen MJ (2014) New approaches to psychiatric diagnostic classification. Neuron 84:564-571.

Perlis RH (2011) Translating biomarkers to clinical practice. Mol Psychiatry 16:1076-1087.

Perrot M, Riviere D, Mangin JF (2011) Cortical sulci recognition and spatial normalization. Med Image Anal 15:529-550.

Poldrack RA, Farah MJ (2015) Progress and challenges in probing the human brain. Nature 526:371-379.

Rayner G, Jackson G, Wilson S (2016) Cognition-related brain networks underpin the symptoms of unipolar depression: evidence from a systematic review. Neurosci Biobehav Rev 61:53-65.

Rosenberg MD, Finn ES, Scheinost D, Papademetris X, Shen X, Constable RT, Chun MM (2016) A neuromarker of sustained attention from whole-brain functional connectivity. Nat Neurosci 19:165-171.

Sakai Y (2015) in The 38th Annual Meeting of the Japan Neuroscience Society (Kobe, Japan, 28-31, July)

Scheinost D, Stoica T, Saksa J, Papademetris X, Constable RT, Pittenger C, Hampson M (2013) Orbitofrontal cortex neurofeedback produces lasting changes in contamination anxiety and resting-state connectivity. Transl Psychiatry 3:e250.

Schnurr PP, Friedman MJ, Engel CC, Foa EB, Shea MT, Chow BK, Resick PA, Thurston V, Orsillo SM, Haug R, Turner C, Bernardy N (2007) Cognitive behavioral therapy for posttraumatic stress disorder in women: a randomized controlled trial. JAMA 297:820-830.

Shibata K, Watanabe T, Sasaki Y, Kawato M (2011) Perceptual learning incepted by decoded fMRI neurofeedback without stimulus presentation. Science 334:1413-1415.

Shibata K, Watanabe T, Kawato M, Sasaki Y (2016) Differential activation patterns in the same brain region led to opposite emotional states. PLoS Biol 14:e1002546.
42

**Figure legends**

**Figure 1:** Distribution of weighted linear summations (WLS) calculated by functional connections. (a) The white and black bars denote the number of typically developing (TD) and autism spectrum disorder (ASD) individuals in the Japanese dataset, respectively. A horizontal axis denotes WLS score. If the WLS score is positive, an individual is classified as having ASD, while a negative WLS score indicates TD. (b) A histogram shows the distribution of WLS scores for the US ABIDE dataset. (c) The density distribution of WLS when applying the ASD classifier to various psychiatric conditions, such as (i) ASD, (ii) SCZ, (iii) ADHD, and (iv) MDD. In each panel, TD/healthy control distribution is gray-colored and ASD distribution is red-colored. The distribution of other psychiatric conditions (i.e., SCZ, ADHD, and MDD) is colored with blue, green, and yellow, respectively. AUC values are based on the classification between each psychiatric condition and TD/healthy control. P-values are obtained by the Benjamini-Hochberg-corrected Kolmogorov-Smirnov test. The TD distribution of WLS at each panel is adjusted to have the same median and s.d. for the visualization purpose. Adapted, with permission, from Figure 1 and 5 in Yahata et al. A small number of abnormal connections predicts adult autism spectrum disorder. Nature Communications, DOI: 10.1038/ncomms11254 (2016).

**Figure 2:** The procedure of decoded neurofeedback (DecNef). During training, participants were instructed to self-regulate brain activity so as to maximize the feedback score. This was represented by, for example, the size of a green disc, which corresponded to the participant's success in inducing a current brain activity pattern as similar as possible to the target brain activity pattern.

**Figure 3:** The procedure of functional connectivity-based neurofeedback (FCNef). During training, participants were instructed to self-regulate brain activity so as to maximize the feedback score. This was represented by, for example, the size of a green disc, which reflected the degree of success in achieving target functional connectivity.

**Figure 4:** Results from three individuals with depression and seven subclinical participants. (a) and (b) show the results of participants with depression. (c) and (d) show the results of participants with subclinical depression. (a) Neurofeedback scores across the four training days. Red bar denotes the mean of neurofeedback scores for all trials. Error bar denotes standard error of the mean. Asterisk shows the statistical significance ($p < 0.001$). (b) Hamilton Depression Rating Scale (HDRS) scores at pre- and post- functional connectivity-based neurofeedback (FCNef). Red

bar denotes the mean of HDRS scores and error bar shows standard error of the mean. (c) Neurofeedback scores in the same format as **a**. Asterisk shows the statistical significance ($p < 0.01$). (d) Scatter plot of the change in the Beck Depression Inventory (BDI) score versus the change of the target rs-fcMRI between post- and pre- neurofeedback. Each dot represents individual data. The line denotes the linear regression of the change of BDI score from the change of the target rs-fcMRI.

**Figure 5:** Neurofeedback-induced change of functional connectivity (FC) toward the neurotypical pattern in a case of adult high-functioning autism spectrum disorder (ASD). The graph shows the feedback scores during the training sessions (blank squares and error bars) and the outputs of the ASD biomarker (Yahata et al., 2016) using the resting-state FC data collected before (i.e., RS-1 and RS-2) and after (i.e., RS-3) the neurofeedback training (x signs). The open circle denotes the mean output of the ASD biomarker across the three rs-fcMRI sessions conducted in a single day. The output of the ASD biomarker ranged between 0 (typical ASD pattern) to 100 (neurotypical pattern). In each training day, there were six runs (filled squares and error bars; except for three runs in the final day), each of which had ten trials. Note that, whereas the outputs of the biomarker had remained close to 0 before the training, the resting state FCs exhibited the neurotypical pattern at least twice out of three scans in the post training, which was acquired three weeks after the training.